\documentclass{article}
\usepackage{jamaica04}
\usepackage{graphicx}
\usepackage{amssymb}
\frompage{000} \topage{000}                                              %|
\def\rightmark{Comprehensive New RHIC II Detector}
\def\leftmark{J.W. Harris et al.}
\newcommand{\pp} {\mbox{$p$$+$$p$}}
\newcommand{\pA} {\mbox{$p$$+$$A$}}
\newcommand{\NN} {\mbox{$A$$+$$A$}}

\newcommand{\pbar} {\mbox{$\overline{p}$}}

\newcommand{\xBJ} {\mbox{x$_{BJ}$}}
\newcommand{\pT} {\mbox{$p_\perp$}}

\newcommand{\snn} {\mbox{$\sqrt{s_{NN}}$}}

\newcommand{\qbar} {\mbox{$\overline{q}$}}

\def\lt{\mbox{$<$}}

\title{A Comprehensive New Detector for Detailed Study of the
QGP, Initial Conditions and Spin Physics at RHIC II\\}
\authors{
{J.W. Harris,$^1$ R. Bellwied,$^2$ N. Smirnov,$^1$ P.
Steinberg,$^3$ B. Surrow$^4$ and T. Ullrich$^3$%
}\\[2.812mm]
{\normalsize \hspace*{-8pt}$^1$ Yale University, New Haven CT, USA 06520-8124\\[0.2ex]
\hspace*{-8pt}$^2$ Wayne State University, Detroit MI, USA 48201\\[0.2ex]
\hspace*{-8pt}$^3$ Brookhaven National Laboratory, Upton NY, USA 11973\\[0.2ex]
\hspace*{-8pt}$^4$ Massachusetts Institute of Technology, Cambridge MA, USA 02139-4307\\[0.2ex]
}}

\abstract{A case is presented for compelling physics at a high
luminosity RHIC II collider. A comprehensive new detector system
is introduced to address this physics. The experimental focus is
on detailed jet tomography of the quark gluon plasma (QGP),
measuring gluon saturation in the nucleus, investigating the color
glass condensate, measuring effects of the QCD vacuum on particle
masses, determining the structure and dynamics within the proton
and possible new phenomena. The physics and detector capabilities
are introduced.}

\PACS{specifications see, e.g.\ {\tt http://www.aip.org/pacs/}}

\begin{document}

\maketitle
\setcounter{page}{1}

\section{Overview}\label{intro}

There are compelling physics questions that can be addressed at a
future, high-luminosity RHIC II complex. These include: What are
the properties of the quark-gluon plasma? To what extent is there
gluon saturation or a color glass condensate in the nucleus at low
Bjorken-x ($\xBJ$)? If present, how does a color glass condensate
evolve into a quark-gluon plasma? What is the chiral structure of
the QCD vacuum and its influence on, or contributions of different
QCD vacuum states to, the masses of particles? What is the
structure and dynamics inside the proton (including spin, possibly
orbital angular momentum) and to what extent is parity violation
significant and important in understanding the proton in the
Standard Model?

The RHIC II complex will be the only QCD facility to have the
capability to address these questions.  In this paper a
comprehensive new detector system is proposed for RHIC II to
address these in an effective way. This detector would utilize
precision tracking and particle identification to large transverse
momentum ($\sim$ 20 GeV/c) in a 1.5 T solenoidal magnetic field,
with electromagnetic and hadronic calorimetry and muon
identification over -3 $< \eta <$ 3 with complete azimuthal
coverage. An in-depth experimental program utilizing the unique
features of this detector system and that of RHIC II is proposed
to answer these compelling physics questions (above) in an era
with heavy ions in the Large Hadron Collider (LHC). Furthermore,
this physics is complementary to the LHC ion program and a future
eRHIC (electron-ion collider) program.

\section{Jet Physics}\label{Jet Physics}

A compelling focus of $\NN$ physics at RHIC II is the use of hard
scattering, in the form of jets and large transverse momentum
(high \pT) particles, to perform a "tomographic" study (with
partons in a multi-parameter space) of the QGP formed at RHIC
\cite{Wang,WangWang,Vitev}.  These studies will be carried out as
a function of geometry (colliding system and impact parameter),
reaction plane, \snn, \pT, rapidity, and particle type. To be able
to accomplish this the detector system should contain full
coverage of electromagnetic and hadronic calorimetry for
measurements of jets and photons, triggering and correlations.
High resolution tracking in a large integral magnetic field and
track-length ($\int$ B $\cdot$ dl) with particle identification up
to large transverse momenta ($\sim$ 20 GeV/c) is essential for
flavor dependence of leading particles and detailed jet
fragmentation studies. High rate data-acquisition and triggering
capabilities are necessary to utilize effectively the high
luminosity for low cross section measurements and photon-,
particle-, and jet-correlations. It should be emphasized that the
flavor dependence in photon-jet, photon-leading hadron,
di-hadrons, and di-jets will be studied as a function of $\xBJ$
and orientation relative to the reaction plane. This complex set
of correlation data will be necessary for detailed determination
of the energy loss mechanism and properties of the QGP.

In order to study the flavor dependence of jet quenching
\cite{Dokshitzer-Kharzeev}, displaced vertices will be used to
identify and trigger on heavy flavor decays. A high $\pT$ electron
in coincidence with a leading hadron, both emanating from a vertex
displaced from the primary $\NN$ reaction vertex, will provide a
trigger for heavy flavor decays. Examples of other specific decay
modes of interest for a displaced vertex, heavy flavor trigger are
B $\rightarrow$ J/$\psi$ + K$^0_s$, and D-mesons through their
hadronic decay modes.

\begin{figure}[htb]
\vspace*{-.2cm}
    \begin{center}
    \includegraphics[width=0.55\textwidth]{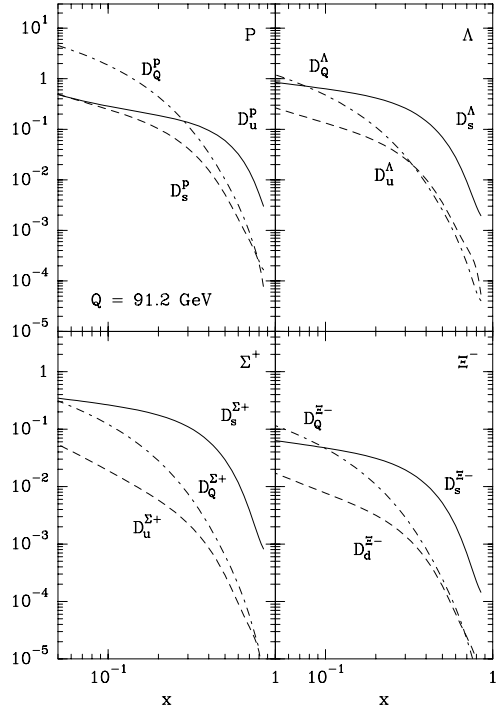}
    \vspace*{-0.6cm}
    \caption[]{Fragmentation functions (D$^{baryon}_{quark}$) for the octet
baryons (p, $\lambda$, $\Sigma$, and $\Xi$) as a function of
$\xBJ$ \cite{Bourrely-Soffer}. Contributions of the light (u,d),
strange (s), and heavy (Q) quarks are denoted by subscripts.}
    \label{fig11}
    \end{center}
\end{figure}

We know that the Higgs field is solely responsible for the mass of
particles in a chirally symmetric medium. The quark condensate
adds a significant part of the mass of each light or strange
particle, when chiral symmetry is broken. The contributions of the
various (u, d, s, heavy) quarks can be determined by measuring the
fragmentation function of each particle in $\pp$ interactions. The
contributions of the light (u,d,), strange (s), and heavy (Q)
quarks to the octet baryons (p, $\lambda$, $\Sigma$, and $\Xi$)
are presented as a function of $\xBJ$ in Fig. \ref{fig11}
\cite{Bourrely-Soffer}. Measurement of these fragmentation
functions requires particle identification of leading particles in
jets at large transverse momentum. Such measurements in $\NN$
collisions will establish how fragmentation functions are modified
by the propagation of the various types of quarks in a dense
medium and should reflect these quark contributions to the
particle masses in the medium. It would be extremely exciting if
these fragmentation functions were to reflect properties of a
chirally restored medium, although this connection has yet to be
established theoretically. In addition to accounting for the
constituent quark masses, the chiral quark condensate is
responsible for inducing transitions between left-handed and
right-handed quarks, $\qbar q$ = $\qbar_{L} q_{R}$ + $\qbar_{R}
q_{L}$. Therefore, helicities of (leading) particles in jets (e.g.
determined by detecting the polarization of leading $\Lambda$
particles) may provide information on parity violation and chiral
symmetry restoration \cite{Kharzeev-Sandweiss}.

In $\pp$ interactions selective kinematic cuts can be implemented
in two-parton scattering to study quark versus gluon jets in
away-side $\pT$ distributions and their particle content. In $\NN$
interactions the ratios of leading anti-particles to leading
particles, such as $\pbar/p$ and K$^-$/K$^+$, at high $\pT$ can be
used to distinguish energy loss differences between gluon and
quark jets \cite{WangWang} in the medium by comparing to results
from $\pp$ and $\pA$ interactions. Although gluon jets dominate at
the highest RHIC and LHC energies, pQCD calculations of jet
production predict that quark jets will become dominant below \snn
$\sim$ 50 GeV at transverse momenta above 8 - 10 GeV/c. The
flexibility and high luminosity of RHIC II will allow detailed
study of both the gluon- and quark-dominated regimes. These
studies require particle identification at large $\pT$.

It is essential to measure the nuclear parton distributions at
low-$\xBJ$ and distinguish $\xBJ$-dependent effects in the initial
stages in order to characterize the evolution of the system from
the initial stages to formation of the QGP. Measurements at large
(forward) rapidities require that faster partons in one projectile
be used to probe soft low-$\xBJ$ partons in the other. Thus, at
large rapidities, we probe aspects of the nuclear wave function
where parton densities are sufficiently large that saturation
phenomena should occur (nuclear shadowing, possibly a color glass
condensate \cite{McLerran,Kharzeev-Levin,KLM}). This also requires
studies of $\pp$ and $\NN$ collisions with instrumentation and
measurements in the forward direction in correlation with those at
mid-rapidity. Jet correlation measurements over a large
pseudorapidity interval with large acceptance in azimuth are
necessary to distinguish the $\xBJ$-dependent evolution of the
collision process and the different properties in the propagation
of light- and heavy-quark jets and gluon jets. All of this is
necessary to understand completely the overall collision process
from a possible color-glass condensate at low-$\xBJ$ in the
colliding nuclei to the formation of the QGP.

\section{Quarkonium Physics}\label{Quarkonium Physics}

The production of quarkonium states in $\pp$, $\pA$, and $\NN$
collisions provides an excellent tool to probe deconfinement in
strongly interacting matter \cite{Matsui}.  To date, studies have
focused on measurements of the charmonium states, which have large
production cross-sections compared to the bottonium states.
Bottonium spectroscopy in $\NN$ has other advantages compared to
charmonium spectroscopy. Bottonium is massive (m $\sim$ 10
GeV/c$^2$) and its decay leptons have large momenta making them
easy to distinguish from background electrons. Furthermore,
interpretation of charmonium suppression is complicated by the
rather large cross-section for absorption by co-moving matter,
while bottonium absorption by hadronic co-movers is expected to be
negligible \cite{Lin}.

Studies of the dependence of the heavy-quark potential on the
in-medium temperature in QCD lattice calculations with dynamical
quarks \cite{Digal} indicate a sequence of melting of the
quarkonium states. Because of its low binding energy the
$\Upsilon$(3S) is expected to dissolve at temperatures below the
critical deconfinement temperature T$_c$. The $\Upsilon$(2S)
dissolves at similar temperatures as the J/$\psi$, while the
$\Upsilon$(1S) remains unsuppressed to temperatures well above
T$_c$. Therefore, a measurement of the yields of the various
bottonium states will shed light on the production (via
$\Upsilon$(1S)) and suppression mechanisms ($\Upsilon$(2S) and
$\Upsilon$(3S)) of quarkonia avoiding many difficulties inherent
in charmonium measurements. These measurements are challenging,
requiring excellent momentum resolution to resolve the bottonium
states and very high rate (luminosity) and trigger capabilities
because of the low production cross-section. These states can be
resolved in the comprehensive new detector system that is
proposed, as will be seen later in this document.

Quarkonium production is not understood to a large extent and
requires further detailed investigation. Since the octet
production matrix elements of NRQCD lead to a polarization pattern
different from the color singlet model, a polarization measurement
can provide significant insight into quarkonium production. For
example, quarkonia at large $\pT$ are predicted to be almost
completely transversely polarized, as they result primarily from
gluon fragmentation. At smaller $\pT$ around $\pT \sim$ 5 GeV/c,
the quarkonia are predicted to be produced essentially
unpolarized. Determination of the polarization as a function of
$\pT$ is essential to test the underlying theory. Recently the
measurement of quarkonium polarization in $\NN$ collisions was
suggested to be an indication of QGP formation \cite{Ioffe}. Since
gluon fragmentation begins to dominate quarkonium production at
$\pT$ $>$ 5 GeV/c , the $\pT$ dependence of quarkonium suppression
should provide a better understanding of the transition from
heavy-quark pair propagation to gluon propagation. Furthermore,
the quarkonium mass establishes the degree of virtuality of the
fragmenting gluon, thus the dependence of the energy loss on the
extent to which the gluon is off-shell can be studied. These
measurements all require a large acceptance detector and large
statistics.

At relatively large transverse momentum typical quarkonium
suppression effects such as color screening become negligible, and
any color octet can suffer jet quenching. This quenching can
provide a unique experimental probe for studying energy loss and
color diffusion \cite{Baier}. The relative yields of charmonium
resonances is an experimental tool for studying such phenomena
because each resonance may have a different octet contribution.
Since a variety of competing charmonium production models exist,
it is essential to investigate production mechanisms in $\pp$ and
$\pA$ interactions, all at low x$_F$ (central region) and high
x$_F$ (forward region) \cite{Gavin,Johnson}. Dead cone or other
effects can be important for heavy quark systems
\cite{Dokshitzer-Kharzeev}. Studies of very high $\pT$ charmonium
can provide additional information to resolve these issues.

\section{Forward Physics}\label{Forward Physics}

One of the major recent developments in high energy nuclear
physics is the \emph{color glass condensate (CGC)}
\cite{McLerran,Kharzeev-Levin}. At very low $\xBJ$, gluons can be
coherent over large longitudinal distances. The coherence reduces
the entropy of the final state. This leads to the importance of
studying high $\pT$ processes away from midrapidity. A signature
of the CGC in the forward region is a systematic hardening of the
particle spectrum due to gluon recombination, associated with a
suppression relative to p+p physics \cite{KLM}.  This may already
be visible in recent d + A data taken by BRAHMS at forward
rapidities \cite{Brahms}, although it must still be determined
whether this is a true saturation effect. Particle identification
(PID) whether measured directly or by means of weak decays, will
be important in elucidating the various particle production
mechanisms, which have different sensitivities to the quark and
gluon components of the hadronic wave functions. Saturation
provides a natural theoretical explanation of the phenomenon of
"nuclear shadowing", a depletion of the nuclear structure function
at low $\xBJ$.  This should have effects on many hard physics
observables which depend directly on the gluon structure, e.g.
minijet rates and heavy flavor production, which can be clarified
by comparisons of $\pA$ physics with $\pp$.

One approach to forward physics at RHIC II is to extend
substantially the detector acceptance forward to allow
\emph{unique global measurements} that are unavailable at the LHC.
These include measurement of high $\pT$ particles as a function of
rapidity out to the kinematic limit, extensive PID capabilities,
and acceptance sufficient to measure total energy flow
event-by-event. To take full advantage of physics in the forward
region, one must be able to perform momentum measurements and PID
up to $\pT$ $\sim$ 2-3 GeV/c, a challenge with longitudinal
momenta of 20-30 GeV/c at large rapidities.  It is also important
to understand the process of energy deposition in the collision
process. Thus, both from the global variable and "parton"
perspectives, there are compelling reasons to fully instrument the
forward region.

Important components of understanding the collision dynamics at
RHIC are the physics underlying the longitudinal particle and
energy distributions and collective flow at mid- and
forward-rapidities. This will require extension of high resolution
particle tracking and calorimetry to forward rapidities. It
remains to be determined how charge, flavor and baryon number are
distributed over all of phase space in RHIC collisions.  These
measurements address fundamental concepts like baryon number
conservation, whose dynamics are not fully understood in QCD.

\section{Spin Physics}\label{Spin Physics}

The potentially exciting topics in polarized proton-proton
collisions that are of interest for a new spin physics program at
RHIC II can be divided into four areas: heavy quark production,
jet physics, electroweak physics, and physics beyond the Standard
Model. The strength of a RHIC II polarized $\pp$ program to probe
effectively these rare processes will depend on the capabilities
of this comprehensive new RHIC II detector.

The \emph{production of heavy quarks} is dominated in $\pp$
collisions by gluon-gluon fusion. In a leading order approximation
\cite{Contogouris} heavy-quark production in polarized $\pp$
collisions constrains the underlying gluon polarization.
Next-to-leading-order (NLO) QCD corrections to the production of
heavy flavors \cite{Bojak} are important for reliable predictions.
Charm and bottom production access different regions of $\xBJ$.
Heavy flavor measurements will rely on leptonic decay channels. In
addition, B-tagging via the J/$\psi$ (B $\rightarrow$ J/$\psi$ +
X) through displaced electron vertices provides identification of
open beauty and probes the gluon polarization.

The \emph{unpolarized production of heavy flavors} has attracted
much attention recently, since beauty production has exhibited
significant discrepancies between theory \cite{Asakawa} and the
data reported at HERA \cite{Klimek} and LEP \cite{Acciarri}. This
has led to descriptions of bottom production in terms of physics
beyond the Standard Model (SM). RHIC II could play an important
role in understanding this discrepancy through investigation of
the energy- and spin-dependent charm and bottom production.

The production of \emph{jets in polarized $\pp$ interactions}
probes the gluon polarization \cite{Bunce} in the proton. The
quark-gluon Compton, gluon-gluon and gluon-quark processes are
important to jet production. The addition of hadronic energy
information in the comprehensive detector would allow precision
jet measurements, correlations and triggering, while extending the
RHIC II polarization program to larger jet $\pT$ and
pseudorapidity.

Measurement of the \emph{parity-violating longitudinal single-spin
asymmetry} A$_L$ in W production is a probe of the underlying
polarized quark and anti-quark distributions \cite{Nadolsky}.
Results obtained in polarized DIS experiments suggest that the QCD
sea is significantly polarized \cite{Adams}. It is crucial to
explore W production in polarized $\pp$ collisions to determine if
the polarization of the QCD sea is shared by quarks and
anti-quarks and if there is any flavor dependence. The difference
of the unpolarized anti-quark distributions is nonzero in the
region of small $\xBJ$ \cite{Peng}. This strong breaking of SU(2)
symmetry has several possible explanations in non-perturbative QCD
\cite{Dressler}. The measurement of W production through the
electron/muon decay channels at high $\pT$ provides a clean
separation of the underlying quark distributions. A hermetic
detector system as proposed would allow measurement of the missing
energy in the unobserved final-state neutrino and reconstruction
of the underlying kinematics.

There may be \emph{physics beyond the standard model (SM)} in the
form of new parity-violating interactions. Parity violation arises
within the SM for quark-quark scattering through an interference
of gluon- and Z$^o$-exchange. There are differing predictions for
various scenarios of physics beyond the SM, e.g. contact
interactions and supersymmetric models \cite{Taxil}. A polarized
RHIC II program will place constraints on several models by
selection of a specific region of phase space unconstrained by the
experimental efforts at the Tevatron. RHIC II would therefore be
in a unique position to explore physics beyond the SM, and the
capabilities of a comprehensive new detector benefit these
studies.

The present understanding of the origin of spin asymmetries is
based on the propagation of polarized quarks and gluons in
external color fields of the hadron. One can understand this
physics better if the strength of the color field can be varied by
utilizing nuclei in polarized p + A collisions.

\begin{figure}[htb]
\vspace*{-.2cm}
%    \begin{center}
    \includegraphics[width=0.9\textwidth]{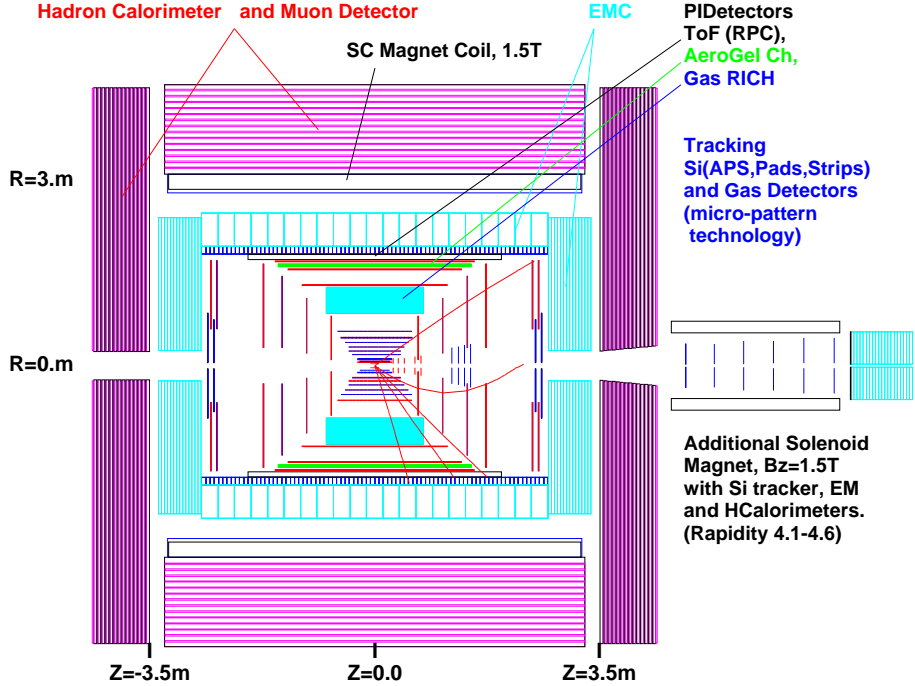}
    \vspace*{-0.5cm}
    \caption[]{Diagram of the new detector using the SLD magnet.}
    \label{fig1}
%    \end{center}
\end{figure}

\section{Comprehensive New Detector and Performance}\label{New Detector}

The requirements for a new RHIC II detector are well defined by
the physics topics discussed in the previous sections and are
quite stringent. The primary requirements are: 1) excellent
charged particle momentum resolution to $\pT$ = 40 GeV/c in the
central rapidity region; 2) complete hadronic and electromagnetic
calorimetry over a large phase space ($\sim$ 4$\pi$); 3) particle
identification out to large $\pT$ ($\sim$ 20 - 30 GeV/c) including
hadron ($\pi$,K,p) and lepton (e/h, $\pi$/h) separation in the
central and forward region; 4) high rate detectors, data
acquisition, and trigger capabilities.

In particular, we expect to procure an existent high field magnet,
muon tracking, and a large amount of electromagnetic and hadronic
calorimetry. As a proof of principle, we employ the SLD magnet,
which is available, with additional super-conducting coils to
increase the field strength to B = 1.5 T. The inner radius of the
magnet is 2.8 m, with an inside length of 6 m. The large magnet
diameter is necessary for tracking and particle identification
detectors and electromagnetic calorimetry (EMC). This provides a
bending power of 3.0 T$\cdot$m over a tracking volume of radius 2
m. The SLD magnet comes with hadronic calorimetry and muon
detection embedded in the iron covering -3 $\leq$ $\eta$ $\leq$ 3.
A possible layout for a RHIC II detector using the SLD magnet is
shown in Fig. \ref{fig1}.

\begin{figure}[htb]
%\vspace*{-.2cm}
    \begin{center}
    \includegraphics[width=0.7\textwidth]{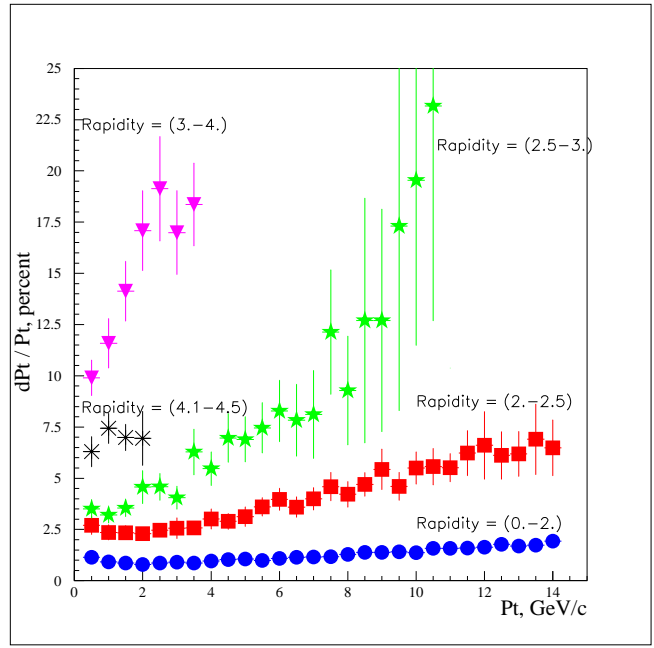}
    \vspace*{-0.5cm}
    \caption[]{Momentum resolution d$\pT$/$\pT$ in percent for
pseudorapidity cuts 0 $\leq \eta \leq$ 2 ($\bullet$), 2  $< \eta
\leq$ 2.5 ($\blacksquare$), 2.5 $< \eta \leq$ 3 ($\ast$), 3 $<
\eta \leq$ 4 ($\circ$), and 4.1 $\leq \eta \leq$ 4.5
($\blacktriangle$).} \label{fig2}
    \end{center}
\end{figure}

A very high resolution vertex detector made from 4 - 5 layers of
thin silicon will be necessary for unambiguous track-seed
determination and for displaced-vertex tagging of b- and c-jets.
With silicon and micro-pattern pad detectors in the SLD magnet we
should reach a momentum resolution of $\delta$$\pT$ /$\pT$ $\sim$
1 $\%$ at 20 GeV/c and $\sim$ 3 $\%$ at 40 GeV/c. The momentum
resolution from full-scale simulations of the detector setup of
Fig. \ref{fig1} are presented in Fig. \ref{fig2}. The two-track
resolution for any charged-particle pair will not exceed 500
$\mu$m.

The requirement to identify all hadrons in a high $\pT$ jet
requires good hadron identification up to momenta of approximately
20 GeV/c. Lepton particle identification will be achieved through
the e/h capabilities in the calorimeters and the muon chambers.
Hadron and lepton particle identification will be achieved through
a combination of dE/dx in the tracking ($\pT$  $\lt$ 1 GeV/c), a
time-of-flight device ($\pT$  $\lt$ 3 GeV/c), and a combination of
two different Aerogel Cherenkov-threshold counters and a RICH
detector with gas radiator  (up to $\pT$ $\sim$ 20 GeV/c). The
time of flight device should be based on resistive plate chambers
with operation in a large magnetic field. For more details on the
comprehensive new detector see \cite{Harris}.

An EMC would be installed directly in front of the SLD magnet
coil, i.e. at 2.8 m radius covering the barrel and in front of the
endcap hadron calorimeter. Different technologies are presently
foreseen for the EMC barrel and endcap sections. A
fine-granularity crystal detector, such as the existing CLEO
crystal calorimeter with its excellent energy resolution, would be
superior and is under consideration. With the high luminosity at
RHIC II and fast detectors there will be sufficient statistics in
the away-side jet and particle spectra for photon-tagged jets out
to a photon $\pT$ $\sim$ 20 GeV/c in this detector.

\begin{figure}[htb]
%\vspace*{-.2cm}
    \begin{center}
    \includegraphics[width=0.75\textwidth]{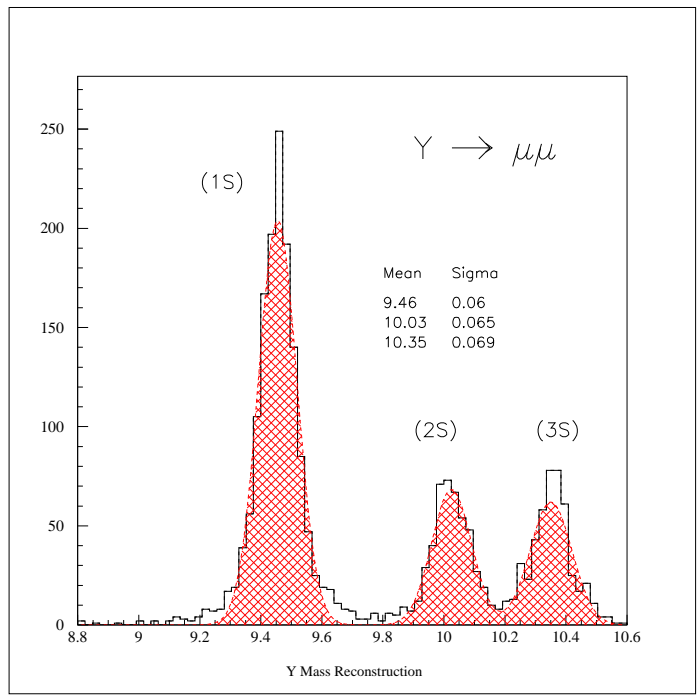}
%\insertplot{nikolai2.ps}
    \vspace*{-0.5cm}
    \caption[]{Mass resolution
        for the $\Upsilon$(1S), $\Upsilon$(2S) and $\Upsilon$(3S)
        states from reconstructed muon tracks in
    simulations. The background is not realistic for $\NN$
    collisions.} \label{fig3}
    \end{center}
\end{figure}

Studies have shown that an energy resolution of better than
10$\%$/$\sqrt(E)$ is required to resolve the quarkonium states
with calorimeter information alone. This clearly shows that the
quarkonia physics can only be fully realized when using the EMC in
combination with high resolution tracking and a muon chamber.
Displayed in Fig. \ref{fig3} are the results of simulations using
the experimental setup of Fig. \ref{fig1}.

The detector envisioned here requires acceptance far exceeding the
current RHIC experiments, both in terms of particle tracking and
calorimetry.  This is driven by requirements of both the
proton-proton program, which needs forward coverage (e.g. to
detect the decay products of vector bosons) as well as the $\pA$
and $\NN$ program, that are concerned both with the dynamics of
nuclear stopping and the details of nuclear shadowing.

With respect to bulk matter measurements in the forward direction,
precise tracking is required to identify any modification in
spectral shape with increasing rapidity. To make heavy flavor and
jet measurements, several approaches can be taken in the forward
region. Beyond $\eta$ = 3 dedicated particle identification and
tracking become complicated. A high precision 'plug' calorimeter
consisting of a crystal-based electromagnetic section and a
hadronic calorimeter based on Si-W could be used in the small
opening of the SLD magnet in the forward direction. Finally, very
forward hadron detectors based on Roman pots instrumented with
scintillating fiber spectrometers can be located at very large
distance (30 - 50 m) from the interaction vertex. These detectors
allow measurement of the forward protons for triggering on
diffractive processes (e.g. pomeron exchange, rapidity gap
measurements). Using a combination of additional strong dipoles
(e.g. RHIC DX design) in the forward direction, tracking modules
(GEM and Si strips) near the beam-pipe, and calorimeter modules
just outside these, one may be able to achieve full coverage for
complete tracking and energy measurements out to very large
pseudo-rapidity.

\section*{Acknowledgement(s)}
The authors wish to thank D. Kharzeev, M. Gyulassy, B. Mueller,
and U. Wiedemann for useful discussions and ideas. This work was
supported by the Office of Science of the US DOE under grant
number DE-FG02-91ER-40609.

\vfill\eject
\end{document}